\documentclass[12pt]{article}
\usepackage{graphicx}
\newcommand{\xxx}[1]{}
\def\[{\left[}
\def\]{\right]}
\def\({\left(}
\def\){\right)}
\newlength{\dummysp}
\newlength{\dummyspl}
\newcommand{\hk}{{\hat k}}
\newcommand{\swZ}{\sin^2 \theta_W (m_Z)}

\newcommand{\beq}{\begin{equation}}
\newcommand{\eeq}{\end{equation}}
\newcommand{\ben}{\begin{enumerate}}
\newcommand{\een}{\end{enumerate}}

\newcommand{\tr}{\mathop{{\hbox{tr} \, }}\nolimits}
\newcommand{\mtxt}[1]{\mathop{\hbox{{\small #1}}}\nolimits}

\newcommand{\ttxt}[1]{\mathop{\hbox{{\tiny #1}}}\nolimits}
\newcommand{\stxt}[1]{\mathop{\hbox{{\scriptsize #1}}}\nolimits}
\newcommand{\bbar}[1]{{\overline{#1}}}
\newcommand{\half}{{1 \over 2}}

\newcommand{\beqa}{\begin{eqnarray}}
\newcommand{\eeqa}{\end{eqnarray}}

\newcommand{\p}{{\partial}}

\newcommand{\vev}[1]{{\langle #1 \rangle}}

\newcommand{\ord}[1]{{{\cal O}(#1)}}

\newcommand{\gappeq}{\mathrel{\rlap {\raise.5ex\hbox{$>$}}
{\lower.5ex\hbox{$\sim$}}}}
\newcommand{\lappeq}{\mathrel{\rlap{\raise.5ex\hbox{$<$}}
{\lower.5ex\hbox{$\sim$}}}}
\newcommand{\myref}[1]{(\ref{#1})}

\newcommand{\ux}{$U(1)_X$}

\newcommand{\LamH}{\Lambda_H}

\newcommand{\LamSB}{{\Lambda_{\ttxt{SUSY}}}}

\newcommand{\LamC}{\Lambda_C}
\newcommand{\LamI}{\Lambda_I}
\newcommand{\LamU}{\Lambda_U}
\newcommand{\uone}{$U(1)$}
\newcommand{\uones}{$U(1)$s}

\newcommand{\icoup}[1]{{\alpha_{#1}^{-1}}}

\newcommand{\db}[1]{\delta b_{#1}}
\newcommand{\dbtw}{{\delta b_2}}
\newcommand{\dbth}{{\delta b_3}}
\newcommand{\dbY}{{\delta b_Y}}

\newcommand{\tHI}{{t_{HI}}}

\newcommand{\ite}[1]{{\it #1 \hspace{2pt}}}

\newcommand{\GGUT}{{G_{\stxt{GUT}}}}

\newcommand{\GSME}{{SU(3)_C \times SU(2)_L \times U(1)_Y}}
\newcommand{\fourth}{{1 \over 4}}
\begin{document}

\begin{titlepage} 

\baselineskip=21pt

\hfill    LBNL-50358

\hfill    UCB-PTH-02/23

\hfill    hep-ph/0205224

\hfill    August 2002

\vspace{10pt}

\begin{center}
{\bf \Large Optical Unification }
\end{center}

\begin{center}
{\sl Joel Giedt${}^*$}
\end{center}

\begin{center}
{\it Department of Physics, University of California, \\
and Theoretical Physics Group, 50A-5101, \\
Lawrence Berkeley National Laboratory, Berkeley,
CA 94720 USA.}\footnote{This work was supported in part by the
Director, Office of Science, Office of High Energy and Nuclear
Physics, Division of High Energy Physics of the U.S. Department of
Energy under Contract DE-AC03-76SF00098 and in part by the National
Science Foundation under grant PHY-00-98840.}
\end{center}

\begin{center}

{\bf Abstract}

\end{center}

We discuss string scale unification facilitated by exotic matter
with masses at intermediate scales, between the observable sector
supersymmetry breaking scale and the string scale.  We point out
a mechanism by which string scale unification may occur while
producing a (lower) virtual unification scale independent
of the location of the intermediate scale and the value of the
string coupling.  The apparent unification obtained by extrapolating
low energy gauge couplings is not accidental when this
mechanism is invoked; virtual unification is robust.

\vfill



\begin{tabbing}

{}~~~~~~~~~\= blah  \kill
${}^*$ E-Mail: {\tt JTGiedt@lbl.gov}

\end{tabbing}

\end{titlepage}

\baselineskip=21pt

\ite{Introduction.}
Four-dimensional weakly coupled heterotic
string theories, such as those constructed from
orbifold compactification \cite{DHVW},
suffer from problems related to unification of
gauge couplings.  Extrapolation of the
low energy (near the Z mass) gauge couplings, assuming
only the spectrum of the Minimal Supersymmetric
Standard Model (MSSM), gives an approximate
unification of the couplings at $\LamU \sim
2 \times 10^{16}$ GeV (see for example \cite{MSSMu}).
On the other hand,
gauge couplings in the weakly coupled
heterotic string should unify at the
string scale $\LamH$, which is roughly
$\LamH \sim 4 \times 10^{17}$ GeV; this is
the infamous problem of a factor of 20.

Various solutions have been suggested for this
difficulty; these have been nicely summarized
by Dienes \cite{Die97}.
One possibility is the following.
Four-dimensional string theories typically
contain exotic states charged under Standard Model (SM) gauge group
$G_{SM} = SU(3)_C \times SU(2)_L \times U(1)_Y$.
These are states beyond those contained 
in the MSSM.  One assumes that some of
these exotics have masses at an
intermediate scale $\LamI$, between the scale of
observable sector supersymmetry breaking $\LamSB$
and the string scale $\LamH$.  The exotic
matter effects the running of the gauge couplings
above $\LamI$ in just the right way to shift
the unification scale up to the string scale.
Recent work on this scenario may be found, for
example, in \cite{Gie02a,Mun01}.

A careful tuning of $\LamI$ is generally
required.  This resolution of the string
unification problem implies that the unification
of gauge couplings in the conventional MSSM picture is
accidental.  It just so happens that the low
energy coupling extrapolate to a unified value.
If the intermediate scale or 
string coupling were perturbed, the apparent unification
at $\LamU$ would disappear.  In general this
unification scenario spoils the significance of
the MSSM $\alpha_3(m_Z)$ versus $\swZ$ prediction \cite{GR99}.

In this paper, we demonstrate a mechanism whereby
string unification no longer implies that the
apparent unification at $\LamU$ is an accident.
In elementary optics a virtual image 
is formed by a diverging thin lens, independent 
of the size and location of the object.  
So too with the mechanism we describe, 
a virtual unification happens regardless of
the intermediate scale (object distance) or
string coupling $\alpha_H$ (object height).
Because of the analogy to the optics of
thin lenses, we refer to the mechanism as
{\it optical unification.}  Optical unification is unlike
the mechanism of {\it mirage unification}
introduced by Ib\'a\~nez in \cite{Ibane99},
since unification is not an illusion;
it really does occur near the string scale
and does not require string threshold
corrections.

\ite{Origin of $\Lambda_I$.}
Quite commonly in the models considered here,
some of the \uone\ factors contained in the string derived
gauge group $G$ are anomalous:
$\tr Q_a \not= 0$.
Redefinitions of the charge generators
allow one to isolate this anomaly such that only one \uone\
is anomalous.  We denote this factor \ux.
The associated anomaly is cancelled by the Green-Schwarz
mechanism \cite{uxr}, which induces
a Fayet-Illiopoulos (FI) term $\xi$ for \ux:
\beq
D_X = \sum_i K_i q^X_i \phi^i + \xi,
\qquad {\xi \over g_H^2} 
= {\tr Q_X \over 192 \pi^2} m_P^2.
\label{1.2}
\eeq
$K_i = \p K / \p \phi^i$, with $K$ the
K\"ahler potential, $q^X_i$ the \ux\ charge of $\phi^i$, 
and $m_P = 1 / \sqrt{8 \pi G} = 2.44 \times 10^{18}$ GeV
is the (reduced) Planck mass.
As discussed below, the (properly normalized) running gauge couplings
$g_a(\mu)$ unify to the heterotic string
coupling $g_H$ at the string scale $\LamH$.
Since the scalar potential
contains a term proportial to $D_X^2$,
some scalar fields generically shift to cancel the FI
term (i.e., $\vev{D_X}=0$) and get large vacuum expectation
values.  This causes several
fields to get effective vector couplings
\beq
W \ni {1 \over m_P^{n-1}} \vev{\phi^1 \cdots \phi^n} A A^c .
\label{ems}
\eeq
Here, $A$ and $A^c$ are conjugate with respect to the
gauge group which survives after spontaneous symmetry
breaking caused by the \ux\ FI term.  
The right-hand side of \myref{ems} is an effective
supersymmetric mass term, which results in masses
$\ord{\vev{\phi^1 \cdots \phi^n} / m_P^{n-1}}$ once one shifts to
the stable vacuum.  This is a possible origin
of an intermediate scale $\LamI$.

Factors of $G$ besides \ux\ are typically
broken when the Xiggses get vevs to
cancel the FI term.  This is an attractive
feature of these models, since it provides
an effective gauge theory with smaller rank.
However, it is typically the case that
\uone\ factors besides the hypercharge $U(1)_Y$
survive below the \ux\ breaking scale.  In those cases where
some of the observable sector fields are charged
under these extra \uones, experimental limits
on $Z'$ boson mediated processes 
require that the unhidden \uones\
be broken somewhere above the electroweak 
symmetry breaking scale.
Thus, we envision the possibility of an
intermediate scale $\LamI$ of gauge
symmetry breaking independent of the \ux\ breaking.

Gaugino condensation
of a condensing group $G_C$ in the hidden sector
gives rise to the condensation scale $\LamC$,
given roughly by $\LamC \sim \LamH \exp (8 \pi^2/ b_C g_H^2)$,
where $b_C$ is the beta function
coefficient of $G_C$ in suitable conventions.
Soft mass terms which
split the supersymmetric partners to Standard Model (SM)
particles are proportional to the gaugino condensate
$\vev{ \lambda \lambda }$.  This operator has
mass dimension three, so one generically expects
the the observable sector supersymmetry breaking
scale $\LamSB$ is given by
\beq
\LamSB \sim \vev{ \lambda \lambda} / m_P^2 \sim \LamC^3 / m_P^2
\eeq
For supersymmetry to protect the gauge hierarchy $m_Z \ll m_P$
between the electroweak scale and the fundamental scale,
one requires $1$ TeV $\lappeq \LamSB \lappeq 100$ TeV.  This implies
$\LamC \sim 10^{13}$ GeV.  
(More precise results may be obtained, for instance,
with the detailed supersymmetry breaking models of Bin\'etruy,
Gaillard and Wu \cite{BGW} as well
as subsequent ellaborations by Gaillard and
Nelson \cite{gn}.)  This gives another dimensionful
parameter which may be used to produce $\Lambda_I$.

\ite{Hypercharge.}
An important feature of Grand Unified Theories (GUTs) \cite{GG74}
is that the \uone\ generator corresponding to
electroweak hypercharge does not have arbitrary normalization,
since the hypercharge generator is embedded into
the Lie algebra of the GUT group $\GGUT \supset \GSME$.
A similar situation holds for the normalization of \uone\
generators in string-derived field theories.
Just as in GUTs, unification of the hypercharge coupling with the
couplings of other factors of the gauge symmetry
group $G$ corresponds to a particular normalization.
This normalization of hypercharge
is often different than the one which appears in $SU(5)$
or $SO(10)$ GUTs.
In our analysis we will work with an arbitrary
hypercharge normalization, conveyed by the constant
$k_Y$.  In an $SU(5)$ GUT, for instance, $k_Y=5/3$.
However, in the semi-realistic heterotic orbifold
constructions, $k_Y$ is typically some other value,
which depends on the linear combination of
\uone\ generators employed to ``manufacture'' the SM hypercharge
generator.  Furthermore, we exploit the fact that
exotic SM-charged states typically carry very
unusual hypercharges; i.e., not those which would
occur in the decomposition of a standard GUT group.
It will turn out that the occurence of the
somewhat bizarre hypercharges
we obtain here fits in nicely with the unusual values
required to realize the optical unification alluded
to above.  (See for example \cite{Gie02a} for a more
complete discussion of these ``stringy'' features.)

\ite{String scale unification.}
It has been known since the
earliest attempts \cite{SS74} to use closed string
theories as unified theories of all fundamental interactions that
$g^2 \sim \kappa^2/\alpha'$,
where $g$ is the gauge coupling, $\kappa$ is the gravitational
coupling and $\alpha'$ is the {\it Regge slope,}
related to the string scale
by $\Lambda_{\stxt{string}} \approx 1/\sqrt{\alpha'}$.
In particular, this relation holds
for the heterotic string~\cite{GHMR85}.
Here, however, $g$ and $\kappa$ are the 
ten-dimensional couplings.  By dimensional reduction 
of the ten-dimensional effective field theory
obtained from the ten-dimensional heterotic string
in the zero slope limit,
this relation may be translated into
a constraint relating the heterotic 
string scale $\LamH$ to the 
four-dimensional Planck mass $m_P$.  
One finds, as expected on dimensional
grounds, $m_P \sim 1/\sqrt{\kappa}$,
where the coefficients which
have been supressed depend on the
size of the six compact dimensions;
similarly, the four-dimensional gauge coupling
satisfies $g_H \sim g$; for details
see Ref.~\cite{DS85}.  We thus obtain
$\LamH \sim g_H m_P$.
Kaplunovsky has made this relation more precise,
including one loop effects from
heavy string states \cite{Kap88}.
Subject to various conventions described
in \cite{Kap88}, including a choice
of the $\overline{\mbox{DR}}$ 
renormalization scheme in
the effective field theory, 
the result is:
\beq
\LamH \approx 0.216 \times g_H m_P
= g_H \times 5.27 \times 10^{17} \mtxt{GeV}.
\label{ssc}
\eeq

In the heterotic orbifolds under
consideration the gauge group $G$ 
has several factors, each of which will
have its own running gauge coupling.
Ginsparg \cite{Gin87} has shown
that the running couplings $g_a^2(\mu)$ unify to a common value $g_H$
at the string scale $\LamH$ according to
\beq
k_a g_a^2(\LamH) = g_H^2, \quad \forall \; a.
\label{hbc}
\eeq
Here, $k_a$ for a nonabelian factor $G_a$ is
the {\it affine} or {\it Kac-Moody level} of
the current algebra in the underlying theory
which is responsible for the gauge symmetry in
the effective field theory.  It is unnecessary
for us to trouble ourselves with a detailed
explanation of this quantity or its string theoretic
origins, since $k_a = 1$ for any nonabelian
factor in the heterotic orbifolds we are considering.
For this reason, these heterotic orbifolds are referred
to as affine-level 1 constructions.
In the case of $G_a$ a \uone\ factor, $k_a$
carries information about the normalization
of the corresponding current in the underlying theory, and
hence the normalization of the charge generator
in the effective field theory.

The important point, which has been emphasized many
times before, is that a gauge coupling
unification prediction is made by the underlying
string theory.  The SM gauge couplings
are known (to varying levels of accuracy), say,
at the Z scale.  Given the
particle content and mass spectrum 
of the theory between the Z scale
and the string scale, one can easily check at the one loop
level whether or not the unification prediction
is approximately consistent with the Z scale boundary values.
To go beyond one loop requires some knowledge of
the other couplings in the theory, and the analysis
becomes much more complicated.  However,
the one loop success is not typically spoiled by
two loop corrections, but rather requires a slight
adjustment of flexible parameters (such as superpartner
masses) which enter the one
loop analysis.

Due to the presence of exotic matter in the models studied here,
we are able to achieve string scale unification.
This sort of unification scenario
has been studied
many times before, for example in \cite{GX92,Far93,Gie02a}.
However, in contrast to \cite{GX92},
we have---as in \cite{Far93}---states which would not appear in
decompositions of standard GUT groups.
The appearance of these states is due to the
Wen-Witten defect in string orbifold models \cite{WW85}.
Using exotics with small hypercharge values,
the $SU(3) \times SU(2)$ running can be
altered to unify at the string scale without
having an overwhelming modification on
the running of the $U(1)_Y$ coupling.

Nonstandard hypercharge normalization has been
studied previously, for example in \cite{Iba93}.  In these analyses,
it was found that {\it lower} values
$k_Y < 5/3$ were preferred if only the
MSSM spectrum is present up to the unification
scale; the preferred values were between $1.4$
to $1.5$.  Unfortunately, in many semi-realistic
orbifold models we are faced with
the opposite effect---a larger than normal
$k_Y$.  This larger value requires a larger correction
to the running from the exotic states,
and has the effect of pushing down the
required mass scale of the exotics \cite{Gie02a}.

\ite{The mechanism.}
With a single intermediate scale $\LamI$,
unification of gauge couplings at the
string scale requires
\beq
k_a 4 \pi \icoup{H} = 4 \pi (\icoup{a}(m_Z) - \Delta_a)
- b_{a} \ln {\LamI^2 \over m_Z^2}
- (b_{a} + \delta b_{a})
\ln {\LamH^2 \over \LamI^2},
\qquad a=Y,2,3.
\label{stu}
\eeq
The notation is conventional, with 
$\alpha_a = g_a^2/ 4 \pi \; (a=H,Y,2,3)$.
The quantities $b_a, \; a=Y,2,3$ are the
beta function coefficients
evaluated for the MSSM spectrum:\footnote{
Here, $C(SU(N))=N$ while $C(U(1))=0$.  For
a fundamental or antifundamental
representation of $SU(N)$ we have $X_a=1/2$
while for hypercharge $X_Y(R) = Y^2(R)$.}
\beq
b_{a} = -3 C(G_a) + \sum_R X_a(R) \quad
\Rightarrow \quad
b_{Y} = 11, \qquad b_{2} = 1,
\qquad b_{3} = -3.
\label{5a}
\eeq
$k_2 = k_3 = 1$, and we leave $k_Y$ unspecified.
$\Delta_a$ are threshold corrections not including
the exotic states that enter the running at $\LamI$.
$\Delta_a$ are immaterial to the mechanism described here---they
drop out of the analysis---therefore there is no
need to estimate or discuss them here.
The quantities $\db{a}$ are the contributions to
the beta function coefficients from exotic states
above $\LamI$:
\beq
\db{3} = \sum_{q_i,q_i^c} \half, \qquad
\db{2} = \sum_{\ell_i,\ell_i^c} \half, \qquad
\db{Y} = \sum_{q_i,q_i^c} (Y_i)^2 + \sum_{\ell_i,\ell_i^c} (Y_i)^2
+ \sum_{s_i,s_i^c} (Y_i)^2 .
\eeq
With respect to $SU(3)_C \times SU(2)_L$
the states $q_i$ are $(3,1)$, $\ell_i$ are $(1,2)$ and $s_i$ are $(1,1)$.
We have vector pairs of each because we assume supersymmetric
masses at the scale $\LamI$.

Virtual unification at $\LamU$ requires
\beq
\hat k_a 4 \pi \icoup{U} = 4 \pi (\icoup{a}(m_Z) - \Delta_a)
- b_{a} \ln {\LamU^2 \over m_Z^2}, \qquad
a=Y,2,3.
\label{msu}
\eeq
Here, $\hat k_2 = \hat k_3 = 1$, while we leave $\hat k_Y = 5/3$,
the usual value. If we substitute \myref{msu} into \myref{stu},
we arrive at the following constraints for
virtual unification at $\LamU$ to occur simultaneously with
real unification at the string scale $\LamH$:
\beq
\hk_a a_U - k_a a_H = b_a (t_H - t_U) + \db{a} \tHI,
\qquad a=Y,2,3,
\label{m5}
\eeq
where for convenience we have defined
$$
a_H = 4 \pi \icoup{H}, \qquad a_U = 4 \pi \icoup{U},
$$
\beq
t_U = \ln (\LamU^2 / m_Z^2), \qquad
t_H = \ln (\LamH^2 / m_Z^2), \qquad
\tHI = \ln (\LamH^2 / \LamI^2) .
\eeq

Generally, Eqs.~\myref{m5} are
inconsistent.  Careful choices of $\delta b_{a}$,
$g_H$ and $\LamI$ must be made.
To have {\it optical unification} we seek
solutions which are independent of
$g_H$ and $\LamI$, except that they correspond
to weakly coupled heterotic string with an
intermediate scale above $m_Z$ and below $\LamH$.
If this can be done, fine-tuning of the intermediate
scale disappears.  We first solve the $a=2,3$ parts of \myref{m5}
for the virtual unification scale
and inverse coupling:
\beq
t_U  =  t_H + \fourth (\dbtw - \dbth) \tHI , \qquad
a_U  =  a_H + \fourth (3 \dbtw + \dbth) \tHI
\label{m8}
\eeq
To have the virtual scale below the string
scale, the first equation in \myref{m8} requires
$\dbtw < \dbth$.
Substituting \myref{m8} into the $a=Y$ part of
\myref{m5} yields the single constraint for
optical unification:
\beq
0 = (k_Y - {5 \over 3}) a_H
+ (\dbY - 4 \dbtw + {7 \over 3} \dbth ) \tHI
\label{m10}
\eeq
Since we require a solution which is independent of
both $a_H$ and $\tHI$, optical unification may
only occur if
\beq
k_Y = 5/3 \qquad {\rm and} \qquad
\dbtw = {7 \over 12} \dbth + \fourth \dbY .
\label{m12}
\eeq
$k_Y = 5/3$ is by no means always possible in four-dimensional
string models.  However, models have been
found where this condition may be satisfied \cite{CFN99,Gie02a}.
In the exotic matter cases considered here, both $\dbtw$ and $\dbth$
will be non-negative integers. 
$\dbY \geq 0$, \myref{m12} and $\dbtw < \dbth$
together imply
\beq
\dbth > \dbtw \geq {7 \over 12} \dbth .
\label{dtr}
\eeq

\ite{GUT-like exotics.}  It is well known that
in the trivial case where the exotics introduced at the
scale $\LamI$ consist entirely of vector pairs of
complete $SU(5)$ representations we retain the
unification prediction.  For example if we have $N$
$5 + \bar 5$ pairs,
\beq
\dbth = \dbtw = N, \qquad \dbY = 5N/3.
\eeq
It is easy to check that this satisfies \myref{m12}.
However, $\dbtw < \dbth$ is not satisfied which from
\myref{m8} is inconsistent with $\LamH > \LamU$.
If we have $N$ $10 + \bbar{10}$ pairs,
\beq
\dbth = \dbtw = 3 N, \qquad \dbY = 5N.
\eeq
While this satisfies \myref{m12},
once again we do not satisfy $\dbtw < \dbth$.

\ite{Minimal example.}
It is easily checked that the minimum
solution to \myref{m12} and \myref{dtr},
consistent with integer values of
$\dbtw$ and $\dbth$, is
\beq
\dbth = 3, \qquad \dbtw = 2, \qquad \dbY = 1.
\label{ato}
\eeq
This can only work if most of the 
$3(3 + \bar 3,1) + 2(1,2+2)$ exotic
states have very small or
vanishing hypercharges.  However,
as we mentioned above it is possible to
have states with very small hypercharge in string-derived models.

For example, if the exotic spectrum which gets a mass
of order $\LamI$ consists of
\beq
3(3 + \bar 3,1,0) + 2(1,2,0) + (1,2,+1/2) + (1,2,-1/2)
\eeq
with respect to $\GSME$, then \myref{ato} is satisfied.
It remains to be seen if this is possible to
achieve in specific models.

\begin{figure}
\begin{center}
\includegraphics[height=6.0in,width=5.0in,angle=90]{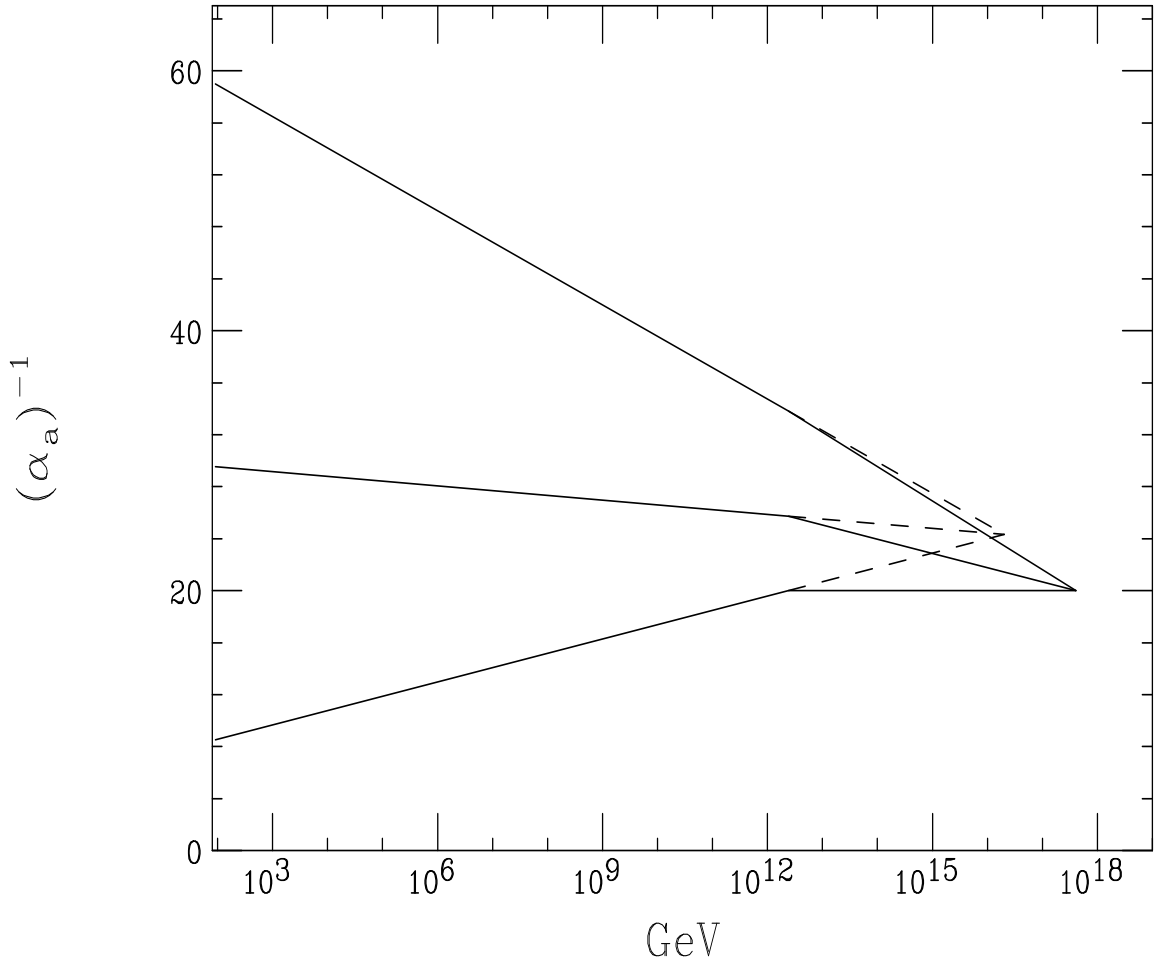}
\end{center}
\caption{Inverse couplings ($\icoup{1} = (3/5) \icoup{Y}$)
as a function of scale for the minimal example.
Solid lines show the actual
running and unification at the string scale.
Dashed lines show the apparent unification.}
\label{fg1}
\end{figure}

In Figure~\ref{fg1} we show the running couplings
in this scheme, assuming negligible Z scale
threshold corrections; i.e., the MSSM superpartners
are assumed to have masses of order the electroweak
scale.  Apparent unification occurs at $\LamU = 2.01
\times 10^{16}$ GeV as can be seen from the figure.
Simultaneously we have string scale unification
at $\LamH =4.08 \times 10^{17}$ GeV and 
\myref{ssc} is satisfied.  The intermediate scale
must be chosen to be $\LamI=2.39 \times 10^{12}$ GeV.

\begin{figure}
\begin{center}
\includegraphics[height=6.0in,width=5.0in,angle=90]{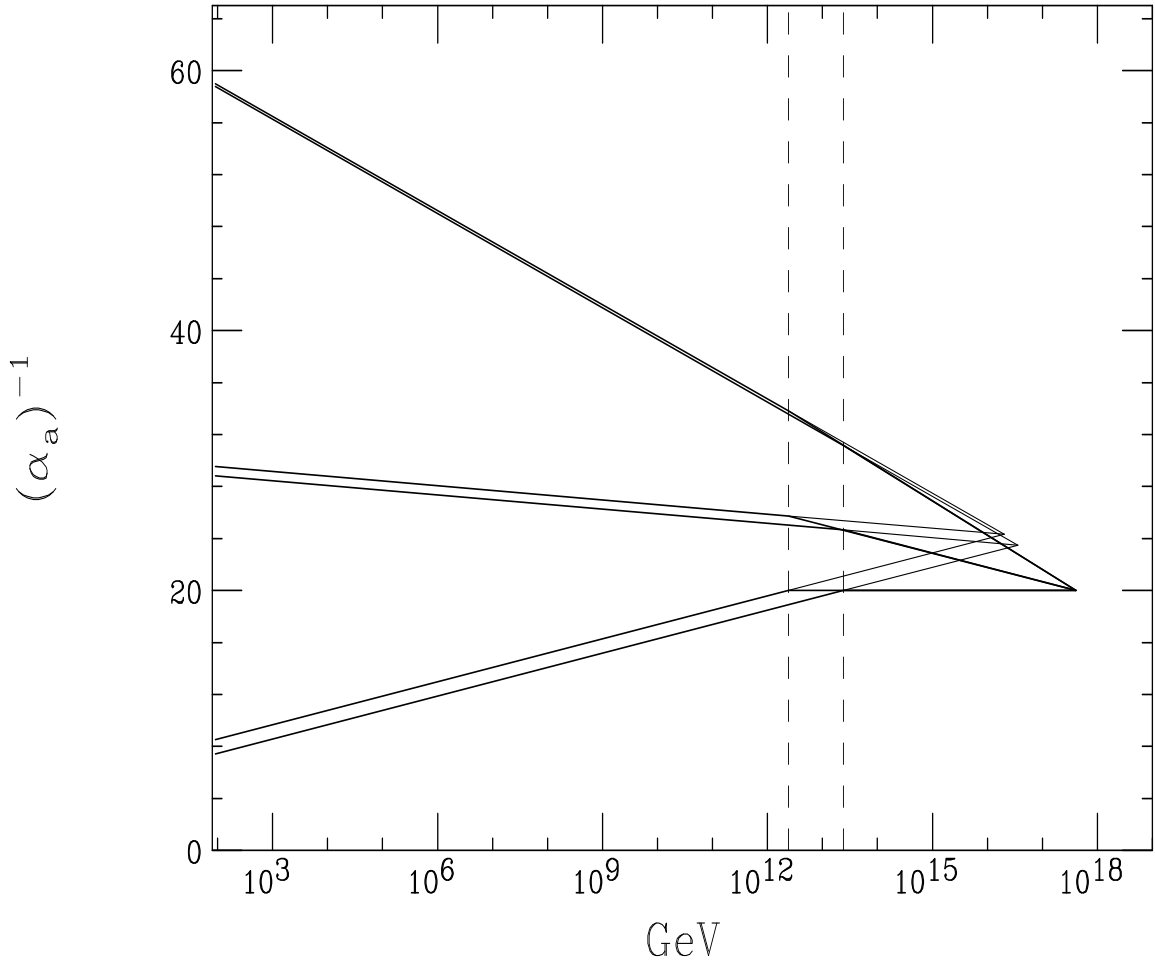}
\end{center}
\caption{Effects of shifting the intermediate
scale upward by a factor of 10, while holding
$\LamH$ and $g_H$ fixed.
Dashed lines show the locations of the two
scales.  Heavy lines show the actual string
scale unification, while apparent unification
is shown with lighter lines.  It can be seen
that apparent unification is shifted to a
slightly higher scale, and the Z scale couplings
must be modified from their experimental values.}
\label{fg2}
\end{figure}

\begin{figure}
\begin{center}
\includegraphics[height=6.0in,width=5.0in,angle=90]{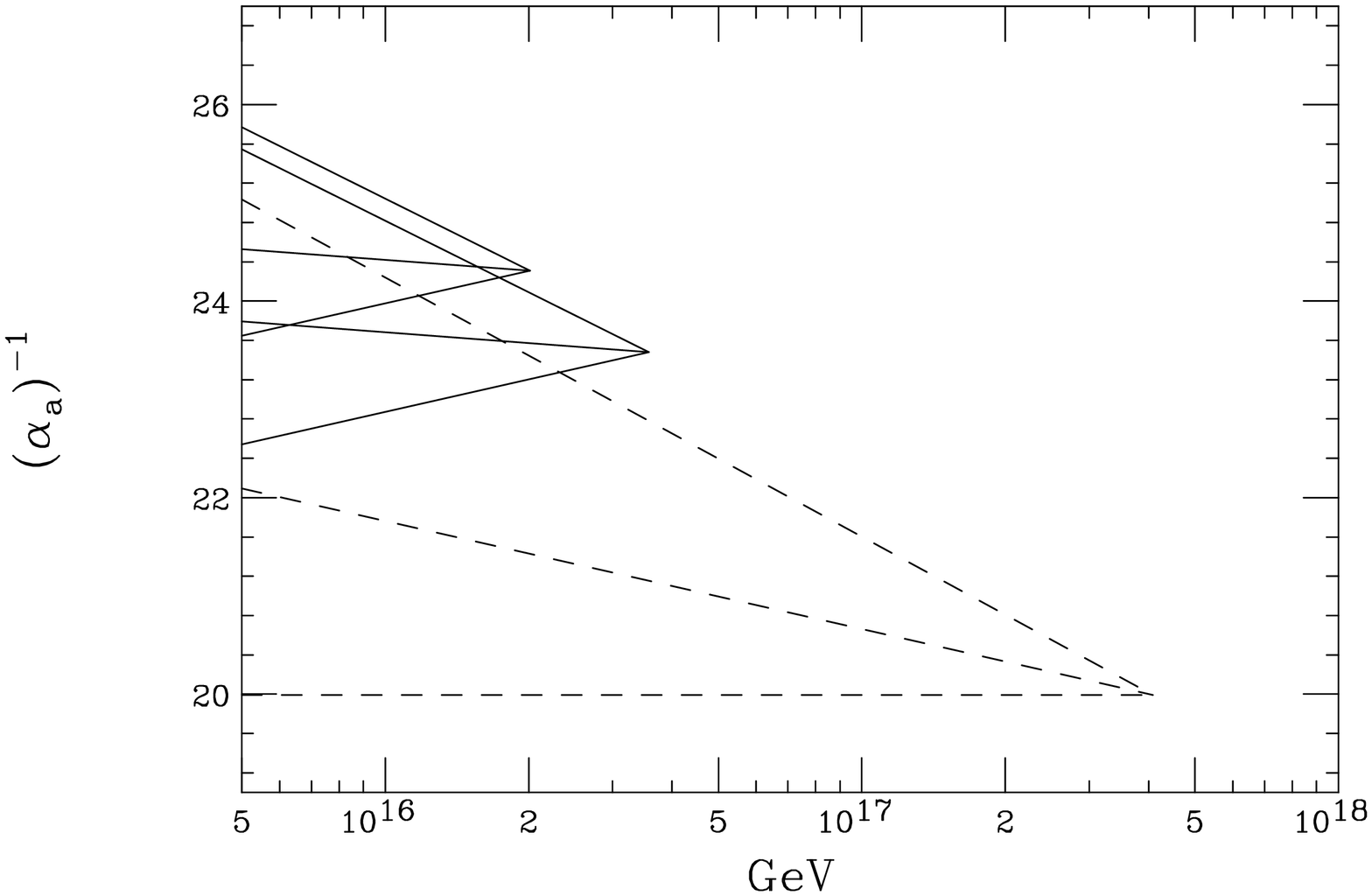}
\end{center}
\caption{Close-up on the effects of shifting the intermediate
scale upward by a factor of 10, while holding
$\LamH$ and $g_H$ fixed.
Dashed lines show the actual string
scale unification, while two versions of
apparent unification
are shown with solid lines.  It can be seen
that apparent unification is preserved.}
\label{fg3}
\end{figure}

If the Z scale couplings are taken as fundamental,
then we must regard $\LamI$ as fine-tuned.  If we shift
$\LamI \to \LamI' = 10 \LamI$, holding the string
scale $\LamH$ and the coupling $g_H$ fixed, then
Figure~\ref{fg2} emerges.  The Z scale couplings must
be shifted away from their experimental values to
accomodate the change in the intermediate scale.
However, as Figure~\ref{fg3} makes clear, apparent
unification is preserved.  Thus Figure~\ref{fg3}
illustrates our main point:  if optical unification
is in force, the {\it qualitative phenomenon} 
of apparent unification is not a 
finely-tuned accident
of a particular choice of intermediate scale.
It is granted, however, that the Z scale couplings and the precise
location of the apparent unification point are
accidents of the choice of intermediate scale.
However, we would like to suggest that these
features are not fundamental, but are a reflection
of the underlying physics at the scales $\LamI$
and $\LamH$.

\ite{Conclusions.}
Clearly a careful tuning of the scale $\LamI$ is
necessary in order to obtain the relation \myref{ssc} and the
observed Z scale boundary values for the running
gauge couplings.  Up to light particle thresholds (MSSM
superpartners, etc.), the quantities 
$\icoup{a},\icoup{U},\LamU$ and $\LamH$
are all determined from
the pair of inputs $(g_H,\LamI)$.  However,
the crucial point is that for {\it any} pair
$(g_H,\LamI)$ we would obtain an apparent unification
at {\it some} scale $\LamU$, so long as \myref{m12}
is satisfied.  That is, the unification obtained
from extrapolation from the Z scale is not accidental.
It is in this sense that the virtual unification
is robust.

In \cite{Gie02a} a class of 175 models was studied
in some detail.  To check whether or not the conditions
for optical unification, Eqs.~\myref{m12}
and \myref{dtr}, can be satisfied requires
an involved analysis of all possible definitions of $Y$
for the 41 models where $k_Y=5/3$ is not excluded.
Moreover, since these models
have many extra $SU(3)_C$ triplets and $SU(2)_L$ doublets,
the number of possible ways of
accomodating the MSSM is vast.
We hope to report the results of such a check
in a future publication.  At this juncture,
we merely point out the mechanism, which has the
nice feature that only the intermediate matter content---and
not the intermediate scale---needs to
be specified in order to {\it naturally} explain
an apparent unification at a scale lower than
the string scale in the context of affine-level 1 
weakly coupled heterotic
string models.  It would be interesting to see how
such a solution to the unification scale problem
might work in semi-realistic heterotic string
models other than those described in \cite{Gie02a},
such as the recent, very promising free
fermionic constructions \cite{CFN99}.

\newpage

\noindent {\bf \Large Acknowledgements}

\vspace{5pt}

\noindent
The author would like to thank Mary K.~Gaillard for
helpful comments, and Christophe Grojean for mentioning
Ref.~\cite{Ibane99}. This work was supported in part by the
Director, Office of Science, Office of High Energy and Nuclear
Physics, Division of High Energy Physics of the U.S. Department of
Energy under Contract DE-AC03-76SF00098 and in part by the National
Science Foundation under grant PHY-00-98840.


\begin{thebibliography}{99}
%
\bibitem{DHVW}
L. Dixon, J. Harvey, C. Vafa and E. Witten,
Nucl. Phys. B 261 (1985) 678;
Nucl. Phys. B 274 (1986) 285.
%
\bibitem{MSSMu}
U. Amaldi, W. de Boer and H. F\"urstenau,
Phys. Lett. B 260 (1991) 447;
J. Ellis, S. Kelly and D. V. Nanopoulos,
Phys. Lett. B 249 (1990) 441.
%
\bibitem{Die97}
K. Dienes, Phys. Rep. 287 (1997) 447\xxx{hep-th/9602045}.
%
\bibitem{Gie02a}
J. Giedt, Ann. of Phys. (N.Y.) 297 (2002) 67\xxx{hep-th/0108244}.
%
\bibitem{Mun01}
C. Mu\~noz, JHEP 0112 (2001) 015\xxx{hep-ph/0110381}.
%
\bibitem{GR99}
D. Ghilencea,
Phys. Lett. B 459 (1999) 540\xxx{hep-ph/9904293};
D. M. Ghilencea and G. G. Ross,
Nucl. Phys. B 606 (2001) 101\xxx{hep-ph/0102306}.
%
\bibitem{Ibane99}
L. E. Ib\'a\~nez, hep-ph/9905349.
%
\bibitem{uxr}
M. B. Green and J. H. Schwarz,
Phys. Lett. B 149 (1984) 117;
M. Dine, N. Seiberg and E. Witten, Nucl. Phys. B 289 (1987) 585;
J. J. Atick, L. Dixon and A. Sen, Nucl. Phys. B 292 (1987) 109; 
M. Dine, I. Ichinose and N. Seiberg, Nucl. Phys. B 293 (1987) 253.
%
\bibitem{BGW}
P. Bin\'etruy, M. K. Gaillard and Y.-Y. Wu,
Nucl. Phys. B 481 (1996) 109; B 493 (1997) 27;
Phys. Lett. B 412 (1997) 288.
%
\bibitem{gn}
M. K. Gaillard and B. Nelson,
Nucl. Phys. B 571 (2000) 3.
%
\bibitem{GG74}
H. Georgi and S. L. Glashow,
Phys. Rev. Lett. 32 (1974) 438.
%
\bibitem{SS74}
J. Scherk and J. H. Schwarz, 
Nucl. Phys. B 81 (1974) 118. 
%
\bibitem{GHMR85}
D. J. Gross, J. A. Harvey, E. Martinec and R. Rohm,
Phys. Rev. Lett. 54 (1985) 502.
%
\bibitem{DS85}
M. Dine and N. Seiberg, 
Phys. Rev. Lett. 55 (1985) 366. 
%
\bibitem{Kap88}
V. S. Kaplunovsky, Nucl. Phys. B 307 (1988) 145; 
Erratum B 382 (1992) 436.
%
\bibitem{Gin87}
P. Ginsparg,
Phys. Lett. B 197 (1987) 139.
%
\bibitem{GX92}
M. K. Gaillard and R. Xiu,
Phys. Lett. B 296 (1992) 71;
S. P. Martin and P. Ramond,
Phys. Rev. D 51 (1995) 6515;
B. C. Allanach and S. F. King,
Nucl. Phys. B 473 (1996) 3.
%
\bibitem{Far93}
A. E. Faraggi,
Phys. Lett. B 302 (1993) 202;
Keith R. Dienes, Alon E. Faraggi,
Phys. Rev. Lett. 75 (1995) 2646\xxx{hep-th/9505018};
Nucl. Phys. B 457 (1995) 409\xxx{hep-th/9505046}.
%
\bibitem{WW85}
X.-G. Wen and E. Witten, 
Nucl. Phys. B 261 (1985) 651.
%
\bibitem{Iba93}
L. E. Ib\'a\~nez, Phys. Lett. B 318 (1993) 73;
K. R. Dienes, A. E. Faraggi and J. March-Russell,
Nucl. Phys. B 467 (1996) 44\xxx{hep-th/9510223}.
%
\bibitem{CFN99}
G. B. Cleaver, A. E. Faraggi and D. V. Nanopoulos,
Phys. Lett. B 455 (1999) 135;
Int. J. Mod. Phys. A 16 (2001) 425;
G. B. Cleaver, A. E. Faraggi, D. V. Nanopoulos, and
J. W. Walker, Mod. Phys. Lett. A 15 (2000) 1191;
Nucl. Phys. B 593 (2001) 471; Nucl. Phys. B 620 (2002)
259\xxx{hep-ph/0104091}.
%
\end{thebibliography}
\end{document}